\documentclass[aps,preprint]{revtex4}%
\usepackage{amsfonts}
\usepackage{amsmath}
\usepackage{amssymb}
\usepackage{graphicx}%
\ifx\pdfoutput\relax\let\pdfoutput=\undefined\fi
\newcount\msipdfoutput
\ifx\pdfoutput\undefined\else
\ifcase\pdfoutput\else
\ifx\paperwidth\undefined\else
\ifdim\paperheight=0pt\relax\else\pdfpageheight\paperheight\fi
\ifdim\paperwidth=0pt\relax\else\pdfpagewidth\paperwidth\fi
\fi\fi\fi
\begin{document}
\preprint{ }
\title{Revisiting the dyonic Majumdar-Papapetrou black holes}
\author{S. Habib Mazharimousavi}
\email{habib.mazhari@emu.edu.tr}
\author{M. Halilsoy}
\email{mustafa.halilsoy@emu.edu.tr}
\affiliation{Department of Physics, Eastern Mediterranean University, G. Magusa, North
Cyprus, Mersin 10 - Turkey.}
\pacs{PACS number}

\begin{abstract}
We extend the Majumdar-Papapetrou (MP) solution of the Einstein-Maxwell (EM)
equations which is implied generally for static electric charge in
non-rotating metrics to encompass equally well magnetic charges. In the
absence of Higgs and non-Abelian gauge fields, 'dyonic' is to be understood in
this simpler sense. Cosmologically this may have far-reaching consequences, to
the extent that existence of multi-magnetic monopole black holes may become a
reality in our universe. Infalling charged particle geodesics may reveal,
through particular integrals, their inner secrets which are screened from our observation.

\end{abstract}
\maketitle

\section{Introduction}

A relatively simple, yet interesting class of Einstein-Maxwell (EM) solution
was given long ago by Majumdar \cite{1} and Papapetrou \cite{2}, which
attracted attention in various contexts including that of multi-black holes.
The isotropic form of the line element with all inclusive metric function,
which determines also the static electric potential makes this solution unique
among EM solutions known to date. The metric function that generates the space
time satisfies the Laplace equation. The linearity of this latter equation
leads automatically to the multi-center solutions at equal ease. Each centre
satisfies all the requirements necessary for black holes and as a matter of
fact the multi-centre solution can be interpreted as a multi-black hole
solution - which is otherwise extremely difficult to obtain analytically. For
this accomplishment we are indebted to the Majumdar-Papapetrou (MP) form of
the metric \cite{3}. A significant extension of the MP solution was to include
time dependence through a cosmological constant \cite{4}. This latter form of
the metric paved the way toward black hole / brane collisions in higher
dimensions \cite{5,6}.

In this paper we wish to contribute to the MP solution by adding magnetic
charge alongside with the electric charge. To our knowledge, MP space time has
been considered so far only with a static electric field described by the
potential $A_{\mu}=\delta_{\mu}^{t}A$, in a diagonal metric. It is known that
inclusion of rotation creates natural magnetic fields from the static electric
charges \cite{7}. Yet, by remaining in the static, non-rotating metric and
adding a magnetic charge to the electric charge - i.e., a dyon - seems escaped
from attentions. Let us note that this should not be confused with solutions
such as Reissner-Nordstr\"{o}m (RN) in which magnetic and electric charges are
treated on equal footing. Our magnetic charge lacks spherical symmetry since
one of the axis (i.e. the z-axis) is singled out which is more apt for
multiple axial superposition. The - dyonic-black holes consist of both
electric and magnetic charges coupled together. Radial geodesic analysis of
electrically charged particles infalling such black holes will exhibit
different behaviors to aid in detection of such magnetic black holes. Under
the light of such magnetic objects the natural question arises: Do such black
holes serve as the storage of magnetic monopoles which are elusive in our
observable universe?

Although it is a matter of formality to extend our results to arbitrary number
of black holes and to higher dimensions, we shall restrict ourselves to
$4-$dimensions and consider the example of $2-$centre black hole as an
example. The problem of horizon smoothness for multi-black holes, and the
issue of stability are two of the problems that we shall not address in this paper.

Organization of the paper is as follows. In Sec. II we present the solution of
EM equations with both electric and magnetic charges. Geodesic analysis
follows for particular boundary /initial conditions. We extend our discussion,
through perturbation, to the case of 2-black holes located along the z-axis.
We complete the paper with Conclusion in Sec. IV.

\section{Integration of the Einstein-Maxwell equations}

We start with the Majumdar-Papapetrou line element in $4-$dimensions
\cite{1,2} given by%
\begin{equation}
ds^{2}=-\frac{1}{\Omega^{2}}dt^{2}+\Omega^{2}\left(  dx^{2}+dy^{2}%
+dz^{2}\right)
\end{equation}
in which $\Omega$ is a function of $x,y,z$ and $t.$ Our electromagnetic
multi-centre potential ansatz is%
\begin{equation}
\mathbf{A}=\frac{\epsilon}{\Omega}dt+%
{\textstyle\sum\limits_{i}}
\frac{P_{i}\left(  z-z_{i}\right)  }{r_{i}\left[  \left(  x-x_{i}\right)
^{2}+\left(  y-y_{i}\right)  ^{2}\right]  }\left[  \left(  x-x_{i}\right)
dy-\left(  y-y_{i}\right)  dx\right]
\end{equation}
where
\[
r_{i}=\sqrt{\left(  x-x_{i}\right)  ^{2}+\left(  y-y_{i}\right)  ^{2}+\left(
z-z_{i}\right)  ^{2}},
\]
$\epsilon$ is a constant such that $0\leq\epsilon\leq1$ and $P_{i}$ stands for
the magnetic charge of the $i^{th}$ black hole. The electric charge $Q_{i}$ of
the $i^{th}$ black hole will be defined below (Eq. 22). The electromagnetic
field two-form is given by%
\begin{equation}%
\begin{tabular}
[c]{l}%
$\mathbf{F}=d\mathbf{A=\epsilon}\left(  \frac{\Omega_{x}}{\Omega^{2}%
}dtdx+\frac{\Omega_{y}}{\Omega^{2}}dtdy+\frac{\Omega_{z}}{\Omega^{2}%
}dtdz\right)  $+$%
{\textstyle\sum\limits_{i}}
\frac{P_{i}}{r_{i}^{3}}\left[  \left(  x-x_{i}\right)  dydz+\left(
y-y_{i}\right)  dzdx+\left(  z-z_{i}\right)  dxdy\right]  $%
\end{tabular}
\end{equation}
with its dual
\begin{equation}
^{\star}\mathbf{F=\epsilon}\left(  \Omega_{x}dydz+\Omega_{y}dzdx+\Omega
_{z}dxdy\right)  +%
{\textstyle\sum\limits_{i}}
\frac{P_{i}}{r_{i}^{3}\Omega^{2}}\left[  \left(  x-x_{i}\right)  dtdx+\left(
y-y_{i}\right)  dtdy+\left(  z-z_{i}\right)  dtdz\right]
\end{equation}
in which $\Omega_{x},\Omega_{y},...$ denote partial derivatives and
$dx^{i}dx^{j}$ implies wedge product. Concerning Maxwell's equations, we have%
\begin{equation}
d\left(  ^{\star}\mathbf{F}\right)  =0
\end{equation}
leading to%
\begin{align}
\nabla^{2}\Omega &  =0,\text{ }\\
\left(
{\textstyle\sum\limits_{i}}
\frac{2P_{i}}{r_{i}^{3}}\right)  \left(  \left(  y-y_{i}\right)  \Omega
_{z}-\left(  z-z_{i}\right)  \Omega_{y}\right)   &  =\epsilon\Omega
_{xt},\nonumber\\
\left(
{\textstyle\sum\limits_{i}}
\frac{2P_{i}}{r_{i}^{3}}\right)  \left(  \left(  z-z_{i}\right)  \Omega
_{x}-\left(  x-x_{i}\right)  \Omega_{z}\right)   &  =\epsilon\Omega
_{yt},\nonumber\\
\left(
{\textstyle\sum\limits_{i}}
\frac{2P_{i}}{r_{i}^{3}}\right)  \left(  \left(  x-x_{i}\right)  \Omega
_{y}-\left(  y-y_{i}\right)  \Omega_{x}\right)   &  =\epsilon\Omega
_{xt},\nonumber
\end{align}
where $\Omega_{xt}=\frac{\partial^{2}\Omega}{\partial x\partial t}$ and so on.
Eq. (6) is the usual Laplace equation whose simplest solution can be written
as%
\begin{equation}
\Omega=\omega\left(  t\right)  +%
{\textstyle\sum\limits_{i}}
\frac{C_{i}}{r_{i}}%
\end{equation}
in which $\omega\left(  t\right)  $ is a function of time and $C_{i}$ are
constants to be identified. A substitution into the rest of Maxwell equations
implies%
\begin{equation}
\frac{\Omega_{x}}{%
{\textstyle\sum\limits_{i}}
\frac{P_{i}}{r_{i}^{3}}\left(  z-z_{i}\right)  }=\frac{\Omega_{y}}{%
{\textstyle\sum\limits_{i}}
\frac{P_{i}}{r_{i}^{3}}\left(  y-y_{i}\right)  }=\frac{\Omega_{z}}{%
{\textstyle\sum\limits_{i}}
\frac{P_{i}}{r_{i}^{3}}\left(  z-z_{i}\right)  },
\end{equation}
which is easily satisfied provided $C_{i}=\lambda P_{i}$ for a constant
$\lambda$. From (3) and (7) we find the field tensor $F_{\mu\nu}$ and the
energy momentum-tensor as
\begin{equation}
T_{\mu}^{\nu}=2F_{\mu\lambda}F^{\nu\lambda}-\frac{1}{2}F\delta_{\mu}^{\nu}%
\end{equation}
in which
\begin{equation}
F=F_{\mu\nu}F^{\mu\nu}=-2\left(  F_{tx}^{2}+F_{ty}^{2}+F_{tz}^{2}\right)
+\frac{2}{\Omega^{4}}\left(  F_{xy}^{2}+F_{xz}^{2}+F_{yz}^{2}\right)  ,
\end{equation}
and
\begin{align}
F_{xy}  &  =%
{\textstyle\sum\limits_{i}}
\frac{P_{i}\left(  z-z_{i}\right)  }{r_{i}^{3}},\text{ \ }F_{xz}=-%
{\textstyle\sum\limits_{i}}
\frac{P_{i}\left(  y-y_{i}\right)  }{r_{i}^{3}},\text{ \ }F_{yz}=%
{\textstyle\sum\limits_{i}}
\frac{P_{i}\left(  x-x_{i}\right)  }{r_{i}^{3}}\\
F_{tx}  &  =\epsilon\frac{\Omega_{x}}{\Omega^{2}},\text{ }F_{ty}=\epsilon
\frac{\Omega_{y}}{\Omega^{2}},\text{ \ \ }F_{tz}=\epsilon\frac{\Omega_{z}%
}{\Omega^{2}}.\nonumber
\end{align}
The non-zero components of $T_{\mu}^{\nu}$ and $G_{\mu}^{\nu}$ are tabulated
in Appendix 1a and 1b, respectively. A solution to the Maxwell Equations (5)
follows once we set (8) to a constant (say $\lambda$) i.e.,
\begin{equation}
\frac{\Omega_{x}}{%
{\textstyle\sum\limits_{i}}
\frac{P_{i}\left(  x-x_{i}\right)  }{r_{i}^{3}}}=\frac{\Omega_{y}}{%
{\textstyle\sum\limits_{i}}
\frac{P_{i}\left(  y-y_{i}\right)  }{r_{i}^{3}}}=\frac{\Omega_{z}}{%
{\textstyle\sum\limits_{i}}
\frac{P_{i}\left(  z-z_{i}\right)  }{r_{i}^{3}}}=\lambda
\end{equation}
or equivalently%
\begin{align}%
{\textstyle\sum\limits_{i}}
\frac{P_{i}\left(  x-x_{i}\right)  }{r_{i}^{3}}  &  =\frac{\Omega_{x}}%
{\lambda},\\%
{\textstyle\sum\limits_{i}}
\frac{P_{i}\left(  y-y_{i}\right)  }{r_{i}^{3}}  &  =\frac{\Omega_{y}}%
{\lambda},\\%
{\textstyle\sum\limits_{i}}
\frac{P_{i}\left(  z-z_{i}\right)  }{r_{i}^{3}}  &  =\frac{\Omega_{z}}%
{\lambda}.
\end{align}
Consequently, the field tensor components become%
\begin{align}
F_{xy}  &  =\frac{\Omega_{z}}{\lambda},F_{xz}=-\frac{\Omega_{y}}{\lambda
},\text{\ }F_{yz}=\frac{\Omega_{x}}{\lambda},\\
F_{tx}  &  =\epsilon\frac{\Omega_{x}}{\Omega^{2}},\text{ }F_{ty}=\epsilon
\frac{\Omega_{y}}{\Omega^{2}},\text{ \ \ }F_{tz}=\epsilon\frac{\Omega_{z}%
}{\Omega^{2}}\nonumber
\end{align}
and as a result the energy-momentum components take the form given in Appendix 1c.

One may also substitute (from the Appendix) into the $tt$ component of the
Einstein's equation with the cosmological constant $\Lambda$ to obtain%
\begin{align}
T_{t}^{t}  &  =G_{t}^{t}+\Lambda\rightarrow\\
-\frac{1}{\Omega^{4}}\left(  \epsilon^{2}+\frac{1}{\lambda^{2}}\right)
\left(  \Omega_{x}^{2}+\Omega_{y}^{2}+\Omega_{z}^{2}\right)   &  =\frac
{1}{\Omega^{4}}\left(  2\Omega\nabla^{2}\Omega-\left(  \mathbf{\nabla}%
\Omega\right)  ^{2}-3\Omega^{4}\Omega_{t}^{2}\right)  +\Lambda\rightarrow
\nonumber\\
-\frac{1}{\Omega^{4}}\left(  \epsilon^{2}+\frac{1}{\lambda^{2}}\right)
\left(  \Omega_{x}^{2}+\Omega_{y}^{2}+\Omega_{z}^{2}\right)   &  =-\frac
{1}{\Omega^{4}}\left(  \mathbf{\nabla}\Omega\right)  ^{2}-3\Omega_{t}%
^{2}+\Lambda.\nonumber
\end{align}
This is satisfied if we make the choices
\begin{equation}
\epsilon^{2}+\frac{1}{\lambda^{2}}=1
\end{equation}
and%
\begin{equation}
3\Omega_{t}^{2}=\Lambda.
\end{equation}
The latter equation implies (from (7)) that%
\begin{equation}
\omega\left(  t\right)  =\pm\sqrt{\frac{\Lambda}{3}}t+C_{0}%
\end{equation}
with an integration constant $C_{0}$ that is disposable with the choice of
origin of time. The rest of the Einstein's equations turn out to be satisfied
all, by virtue of (18) and (19). In conclusion, we obtain the solution as%
\begin{equation}
\Omega=\pm\sqrt{\frac{\Lambda}{3}}t+C_{0}+%
{\textstyle\sum\limits_{i}}
\frac{P_{i}\lambda}{\left\vert \mathbf{r}-\mathbf{r}_{i}\right\vert }%
\end{equation}
which clearly for $\mathbf{r=r}_{i}$ we have the location of the $i^{th}$
black hole with effective charge and mass equal to $\left\vert P_{i}%
\lambda\right\vert .$ With reference to our potential ansatz (2), we observe
that for $\epsilon=0$ we have the pure magnetic charge $P_{i}=m_{i}$ (mass).
To define the electric charge $Q_{i}$ we integrate the Maxwell equation in
accordance with%
\begin{equation}
\oint\vec{E}_{i}.d\vec{A}_{i}=4\pi Q_{i}%
\end{equation}
over the $i^{th}$ sphere to get $Q_{i}=\frac{\epsilon}{\sqrt{1-\epsilon^{2}}%
}P_{i}$. It is clear from this definition that for $\epsilon=1$ we must take
$P_{i}\rightarrow0$ to have a meaningful electric charge, this is indeed the
case as given in the sequel. For a single black hole it is just the extremal
Reissner-Nordstr\"{o}m (RN) black hole solution, as expected. A similar
integral to (22) for the magnetic field reveals also that $P_{i}$ stands for
the magnetic charges. From the balancing gravitational and electromagnetic
force the electric / magnetic charge are proportional to mass in accordance
with
\begin{align}
Q_{i}  &  =m_{i}\epsilon,\\
P_{i}  &  =m_{i}\sqrt{1-\epsilon^{2}},
\end{align}
so that
\begin{equation}
Q_{i}^{2}+P_{i}^{2}=m_{i}^{2}.
\end{equation}

\section{Geodesic equation of an electrically charged test particle}

In this section we seek for a solution to the geodesic equations of a test
charge inside the field of a single static black hole located at the origin
and for simplicity we shall assume $\Lambda=0,$ $C_{0}=1$. The line element is
given by
\begin{equation}
ds^{2}=-\frac{1}{\Omega^{2}}dt^{2}+\Omega^{2}\left(  dr^{2}+r^{2}\left(
d\theta^{2}+\sin^{2}\theta\text{ }d\phi^{2}\right)  \right)  ,
\end{equation}
where%
\begin{equation}
\Omega=1+\frac{m}{r}%
\end{equation}
and $m=$ $\lambda P=Q/\epsilon$. The Lagrangian for a test particle with
electric charge $q$ and unit mass is%
\begin{equation}
\mathcal{L}=-\frac{\dot{t}^{2}}{2\Omega^{2}}+\frac{\Omega^{2}}{2}\left[
\dot{r}^{2}+r^{2}\left(  \dot{\theta}^{2}+\sin^{2}\theta\text{ }\dot{\phi}%
^{2}\right)  \right]  +\frac{q\epsilon}{\Omega}\dot{t}+qP\cos\theta\text{
}\dot{\phi},
\end{equation}
in which a 'dot' stands for derivative with respect to the proper time $\tau.$
This Lagrangian implies the following equations, and first integrals%
\begin{gather}
-\frac{\dot{t}}{\Omega^{2}}+\frac{q\epsilon}{\Omega}=\alpha_{0},\\
\Omega^{2}r^{2}\sin^{2}\theta\text{ }\dot{\phi}+qP\cos\theta=\beta
_{0},\nonumber\\
\ddot{r}+\frac{\Omega^{\prime}}{\Omega}\dot{r}^{2}=-\frac{\alpha_{0}%
\Omega^{\prime}}{\Omega^{2}}\left(  q\epsilon-\alpha_{0}\Omega\right)
+\frac{r}{\Omega}\dot{\theta}^{2}+\frac{\left(  qP\right)  ^{2}}{r^{3}%
\Omega^{5}\sin^{2}\theta}\left(  \beta-\cos\theta\right)  ^{2},\text{
\ \ \ \ \ \ \ }\nonumber\\
\frac{d}{d\tau}\left(  r^{2}\Omega^{2}\dot{\theta}\right)  =\frac{\left(
qP\right)  ^{2}}{r^{2}\Omega^{2}\sin^{3}\theta}\left(  \beta-\cos
\theta\right)  \left(  \beta\cos\theta-1\right)  ,\nonumber\\
\left(  \Omega^{\prime}=\frac{d\Omega}{dr}\right)  ,\nonumber
\end{gather}
where $\alpha_{0}$ and $\beta_{0}$ are two integration constants related to
energy, angular momentum and the constant $\beta$ is defined by \ $\beta
=\frac{\beta_{0}}{qP}.$

We start with the $\theta$ equation, by setting $\theta=\theta_{0}$. This
leads to two different cases:

\subsection{$\beta=\cos\theta_{0},$ ($0<\theta_{0}<\frac{\pi}{2}$)}

By taking $\beta=\cos\theta_{0}$ and $\theta=\theta_{0}$ one easily finds
$\theta$ equation is satisfied and $\phi$ equation requires either $\theta
_{0}=0$ or $\dot{\phi}=0.$ Here we exclude the case of $\theta_{0}=0$ and
accept $\dot{\phi}=0.$ The $r$ \ equation reduces to%
\begin{equation}
\ddot{r}+\frac{\Omega^{\prime}}{\Omega}\dot{r}^{2}=-\frac{\alpha_{0}%
\Omega^{\prime}}{\Omega^{2}}\left(  q\epsilon-\alpha_{0}\Omega\right)  .
\end{equation}
The latter equation yields the following non-linear differential equations.

\subsubsection{Case of $\alpha_{0}=0$}

A specific analytical solution can be found by setting $\alpha_{0}=0.$ This
choice leads to
\begin{equation}
\ddot{r}+\frac{\Omega^{\prime}}{\Omega}\dot{r}^{2}=0
\end{equation}
which reveals
\begin{equation}
r\left(  \tau\right)  =m\text{ LambertW}\left(  e^{A\tau+B}\right)
\end{equation}
in terms of the LambertW$\left(  .\right)  $ function \cite{8}. The constants
$\mathcal{A}$ and $\mathcal{B}$ can be fixed so that $r\left(  \tau\right)
=0$ is reached in a finite proper time. The ordinary time $t$ is also
expressed in terms of the $\tau$ by
\begin{equation}
t\left(  \tau\right)  =\frac{q\epsilon}{A}\left(  \text{LambertW}\left(
e^{A\tau+B}\right)  -\frac{1}{\text{LambertW}\left(  e^{A\tau+B}\right)
}+2\ln(\text{LambertW}\left(  e^{A\tau+B}\right)  )\right)  +C
\end{equation}
in which $C$ is another integration constant. In terms of the coordinate time
$t$, $x(t)$ satisfies the differential equation%
\begin{equation}
x\left(  1+x\right)  \frac{d^{2}x}{dt^{2}}-2\left(  \frac{dx}{dt}\right)
^{2}=0
\end{equation}
which can be studied numerically.

\subsubsection{Case of $\alpha_{0}\neq0$}

For the general case of $\alpha_{0}\neq0,$ we introduce the new parameters
\begin{equation}
r=mx,\text{ \ \ }q\epsilon=m\left(  \mathcal{B+A}\right)  ,\text{ \ \ }%
\alpha_{0}=m\mathcal{A},
\end{equation}
into the Eq. (30) to get%
\begin{gather}
\ddot{x}+\frac{\Omega^{\prime}}{\Omega}\dot{x}^{2}=-\frac{\Omega^{\prime}%
}{\Omega^{2}}\mathcal{A}\left(  \mathcal{B+A}-\mathcal{A}\Omega\right)
\rightarrow\nonumber\\
x\left(  x+1\right)  ^{2}\ddot{x}-\left(  x+1\right)  \dot{x}^{2}%
=\mathcal{A}\left(  \mathcal{B}x-\mathcal{A}\right)  .
\end{gather}
With particular boundary conditions we plot $x\left(  \tau\right)  $ in Fig.
1a directly from this differential equation. Now, in order to obtain a
particular solution we introduce the ansatz (from the analogy of a velocity
dependent potential)
\begin{equation}
\dot{x}^{2}=%
{\textstyle\sum\limits_{k=0}^{\infty}}
a_{k}x^{k}%
\end{equation}
which leads to%
\begin{equation}
\ddot{x}=%
{\textstyle\sum\limits_{k=1}^{\infty}}
\frac{1}{2}ka_{k}x^{k-1}.
\end{equation}
By substitution into (36) we get%
\begin{equation}
a_{0}=\mathcal{A}^{2},\text{ \ \ }a_{1}=-2\mathcal{A}\left(  \mathcal{A+B}%
\right)  ,\text{ \ \ }a_{k}=-\left(  2a_{k-1}+a_{k-2}\right)  =\left(
-1\right)  ^{k+1}\left[  \left(  k-1\right)  a_{0}+ka_{1}\right]  ,
\end{equation}
which imply%
\begin{gather}
\dot{x}^{2}=%
{\textstyle\sum\limits_{k=0}^{\infty}}
\left(  -1\right)  ^{k+1}\left[  \left(  k-1\right)  a_{0}+ka_{1}\right]
x^{k}=a_{0}%
{\textstyle\sum\limits_{k=0}^{\infty}}
\left(  -1\right)  ^{k+1}\left(  k-1\right)  x^{k}+a_{1}%
{\textstyle\sum\limits_{k=0}^{\infty}}
\left(  -1\right)  ^{k+1}kx^{k}=\\
a_{0}\left(  1-\left(  \frac{x}{1+x}\right)  ^{2}\right)  +a_{1}\left(
\frac{x}{\left(  1+x\right)  ^{2}}\right)  =\frac{\left[  \left(  1+2x\right)
a_{0}+xa_{1}\right]  }{\left(  1+x\right)  ^{2}}.\nonumber
\end{gather}
This easily gives%
\begin{equation}
\dot{x}=\frac{\pm1}{\left(  1+x\right)  }\sqrt{\mathcal{A}\left(
\mathcal{A}-2\mathcal{B}x\right)  }%
\end{equation}
and therefore%
\begin{align}
\tau+C_{1} &  =\frac{\pm1}{3\mathcal{AB}^{2}}\left(  \mathcal{A+B}\left(
3+x\right)  \right)  \sqrt{\mathcal{A}\left(  \mathcal{A}-2\mathcal{B}%
x\right)  }\\
\text{(}C_{1} &  =\text{constant).}%
\end{align}
This gives the relation between the proper time and the position of particle
for any value of $\alpha_{0}.$ Next, by using the $t$ component of the
geodesic equation we find
\begin{equation}
\left(  2q\epsilon-3\alpha_{0}\Omega\right)  \Omega^{\prime}\left(  \frac
{dr}{dt}\right)  ^{2}+\Omega\left(  q\epsilon-\alpha_{0}\Omega\right)
\frac{d^{2}r}{dt^{2}}=-\alpha_{0}\frac{\Omega^{\prime}}{\Omega^{3}},
\end{equation}
where $\Omega$ is still given by (27). Here also we rescale our variables as
\begin{equation}
r=mx,\text{ \ \ \ }q\epsilon=m\left(  \mathcal{B+A}\right)  ,\text{
\ \ }\alpha_{0}=m\mathcal{A},\text{ \ }t=m\tilde{t}%
\end{equation}
to get%
\begin{equation}
x\left(  1+x\right)  ^{4}\left(  \mathcal{B}x-\mathcal{A}\right)  \frac
{d^{2}x}{d\tilde{t}^{2}}-\left(  1+x\right)  ^{3}\left(  \left(
2\mathcal{B}-\mathcal{A}\right)  x-3\mathcal{A}\right)  \left(  \frac
{dx}{d\tilde{t}}\right)  ^{2}=\mathcal{A}x^{4}%
\end{equation}
which has the exact solution
\begin{align}
\pm\tilde{t}+C_{2} &  =\frac{1}{3\sqrt{\mathcal{A}}}\sqrt{\mathcal{A}%
-2\mathcal{B}x}\left(  \frac{3}{x}-6-x+\frac{2\mathcal{A}}{\mathcal{B}%
}\right)  +2\ln\left\vert \frac{\sqrt{\mathcal{A}}+\sqrt{\mathcal{A}%
-2\mathcal{B}x}}{\sqrt{\mathcal{A}}-\sqrt{\mathcal{A}-2\mathcal{B}x}%
}\right\vert ,\\
(\mathcal{A} &  \neq0),\text{ \ \ }(\mathcal{A}>2\mathcal{B}x),\text{
\ \ (}C_{2}=\text{constant).}.\nonumber
\end{align}
We can easily observe that for $x\rightarrow0,$ $\tilde{t}\rightarrow\infty$
as expected for a distant observer; Fig. 1b reveals this fact.

\subsubsection{The case of pure magnetic charge}

The case of pure magnetic charge can be obtained by setting $\epsilon=0,$ or
equivalently $\mathcal{B}=-\mathcal{A}.$ This leads to
\begin{equation}
a_{0}=\mathcal{A}^{2},\text{ \ \ }a_{1}=0,
\end{equation}
and therefore%
\begin{align}
\dot{x}  &  =\frac{\pm\left\vert \mathcal{A}\right\vert }{1+x}\sqrt{1+2x},\\
\tau+C_{3}  &  =\frac{\mp1}{3\left\vert \mathcal{A}\right\vert }\left(
2+x\right)  \sqrt{1+2x}\\
\text{(}C_{3}  &  =\text{constant).}%
\end{align}
The latter equation leads to%
\begin{equation}
x\left(  \tau\right)  =-\frac{1}{2}\left(  \sqrt[3]{3\sigma+\sqrt{1+9\sigma}%
}-\frac{1}{\sqrt[3]{3\sigma+\sqrt{1+9\sigma}}}\right)
\end{equation}
in which%
\begin{equation}
\sigma=\pm\left\vert \mathcal{A}\right\vert \tau+C_{3}.
\end{equation}
Eq. (45) becomes now%
\begin{equation}
-x\left(  1+x\right)  ^{5}\frac{d^{2}x}{d\tilde{t}^{2}}+3\left(  1+x\right)
^{4}\left(  \frac{dx}{d\tilde{t}}\right)  ^{2}=x^{4}%
\end{equation}
with exact solution
\begin{align}
\pm\tilde{t}+C_{4}  &  =\frac{1}{3}\sqrt{1+2x}\left(  \frac{3}{x}-x-8\right)
+2\ln\left\vert \frac{\sqrt{1+2x}+1}{\sqrt{1+2x}-1}\right\vert ,\\
\text{(}C_{4}  &  =\text{constant).}%
\end{align}

\subsubsection{The case of pure electrically charged black hole}

By choosing $\epsilon=1,$ $P=0$ in the Lagrangian (28) we obtain a reduced set
of geodesics equations. The $\dot{\theta}=0$ case implies automatically that
$\theta=\frac{\pi}{2}$ and $\dot{\phi}=0=\beta_{0}$. This is nothing but same
as (30) with the additional condition of $\epsilon=1,$ and the resulting
geodesics motion obtained above. Thus, in class-A geodesics, $P=0$ case
doesn't show a significant difference from the $P\neq0$ case.

\subsection{$\beta=1/\cos\theta_{0},$ ($0<\theta_{0}<\frac{\pi}{2}$)}

After setting $\theta=\theta_{0},$ in order to solve $\theta$ equation one can
also choose $\beta=1/\cos\theta_{0}.$ This choice in $\phi$ equation leads%
\begin{equation}
\text{ }\dot{\phi}=\frac{qP}{\cos\theta_{0}}\frac{1}{\Omega^{2}r^{2}},
\end{equation}
and $r=mx$ equation reads%
\begin{equation}
\ddot{r}+\frac{\Omega^{\prime}}{\Omega}\dot{r}^{2}=-\frac{\alpha_{0}%
\Omega^{\prime}}{\Omega^{2}}\left(  q\epsilon-\alpha_{0}\Omega\right)
+\frac{\left(  qP\right)  ^{2}}{r^{3}\Omega^{5}}\tan^{2}\theta_{0}.
\end{equation}
This choice does not change the $t$ equation. Now we use the same change of
variables (35{}) together with
\begin{equation}
q=m\tilde{q},P=m\tilde{P},\left(  \tilde{q}\tilde{P}\right)  ^{2}\tan
^{2}\theta_{0}=\mathcal{C}^{2}%
\end{equation}
under which the $r$ equation takes the form%
\begin{equation}
\ddot{x}+\frac{\Omega^{\prime}}{\Omega}\dot{x}^{2}=-\frac{\Omega^{\prime}%
}{\Omega^{2}}\mathcal{A}\left(  \mathcal{B}-\mathcal{A}\Omega\right)
+\frac{\mathcal{C}^{2}}{x^{3}\Omega^{5}}%
\end{equation}
where $\Omega=1+\frac{1}{x},$ and $\Omega^{\prime}=\partial_{x}\Omega.$ It is
observed that the last term on the right hand side is a direct contribution of
the magnetic charge with marked distinction from the pure electrically charged
black hole case. We note that by some manipulation on the $\phi$ equation, one
gets%
\begin{equation}
\dot{\phi}=\frac{\tilde{q}\tilde{P}}{\cos\theta_{0}}\frac{1}{\Omega^{2}x^{2}%
}=\frac{\mathcal{D}}{\Omega^{2}x^{2}}%
\end{equation}
where $\mathcal{D}=$ $\frac{\tilde{q}\tilde{P}}{\cos\theta_{0}}.$ By
transforming the independent variable from the proper time $\tau$ to the
azimuthal angle $\phi$ the orbit equation takes the form
\begin{equation}
x^{\prime\prime}-\frac{\left(  1+2x\right)  }{x^{2}}\frac{x^{\prime2}}{\Omega
}=\frac{x^{2}\mathcal{A}}{\mathcal{D}^{2}}\left(  \mathcal{B}-\mathcal{A}%
\Omega\right)  +\frac{\mathcal{C}^{2}}{\mathcal{D}^{2}}\frac{x}{\Omega},
\end{equation}
or equivalently
\begin{equation}
\frac{d^{2}x\left(  \phi\right)  }{d\phi^{2}}-\frac{\left(  1+2x\right)
}{x\left(  1+x\right)  }\left(  \frac{dx\left(  \phi\right)  }{d\phi}\right)
^{2}=\frac{x^{2}\mathcal{A}}{\left(  1+x\right)  \mathcal{D}^{2}}\left(
\mathcal{B}x-\mathcal{A}\left(  1+x\right)  \right)  +\frac{\mathcal{C}^{2}%
}{\mathcal{D}^{2}}\frac{x^{2}}{\left(  1+x\right)  }.
\end{equation}
Fig. 2 gives a numerical plot of $x\left(  \phi\right)  ,$ under the boundary
conditions $\left.  x\left(  \phi\right)  \right\vert _{\phi=0}=1$ and
$\left.  \frac{dx\left(  \phi\right)  }{d\phi}\right\vert _{\phi=0}=0.$

\subsection{Generalization to two-centre black holes}

In this section we try to extend the result found for single black hole to
double-black hole system. To do so we consider two identical black holes at
$\left(  0,0,h\right)  $ and $\left(  0,0,-h\right)  ,$ and the test particle
is placed at a distance, far from the black holes such that one can write the
metric function, up to the third order, as%
\begin{equation}
\Omega=1+\frac{2m}{r}+\frac{mh^{2}\left(  3\cos^{2}\theta-1\right)  }{r^{3}%
}+O\left(  \frac{h}{r}\right)  ^{4}.
\end{equation}
It is easily seen that for $h\rightarrow0$ the metric goes to the extremal RN
black hole with both electric and magnetic charges and total mass $2m,$ as it
should. The Lagrangian of the system, up to the same order of approximation,
from the potential (2) can be written as
\begin{equation}
\mathcal{L}=-\frac{\dot{t}^{2}}{2\Omega^{2}}+\frac{\Omega^{2}}{2}\left[
\dot{r}^{2}+r^{2}\left(  \dot{\theta}^{2}+\sin^{2}\theta\text{ }\dot{\phi}%
^{2}\right)  \right]  +\frac{q\epsilon}{\Omega}\dot{t}+qP\left[  \cos
\theta\left(  2-\frac{3h^{2}\sin^{2}\theta}{r^{2}}\right)  +O\left(  \frac
{h}{r}\right)  ^{4}\right]  \text{ }\dot{\phi}.
\end{equation}
This leads to the following geodesic equations (with integration constant
$\alpha_{0}$ and $\beta_{0}$)
\begin{gather}
\dot{t}=\left(  q\epsilon-\alpha_{0}\Omega\right)  \Omega,\\
\Omega^{2}r^{2}\sin^{2}\theta\text{ }\dot{\phi}+qP\left[  \cos\theta\left(
2-\frac{3h^{2}\sin^{2}\theta}{r^{2}}\right)  +O\left(  \frac{h}{r}\right)
^{4}\right]  =\beta_{0},\\
\left(  \ddot{r}+2\frac{\Omega_{r}}{\Omega}\dot{r}^{2}\right)  =-\left(
q\epsilon-\alpha_{0}\Omega\right)  \frac{\Omega_{r}\alpha_{0}}{\Omega^{2}%
}+\frac{\Omega_{r}}{\Omega}\left[  \dot{r}^{2}+r^{2}\left(  \dot{\theta}%
^{2}+\sin^{2}\theta\text{ }\dot{\phi}^{2}\right)  \right]  +\nonumber\\
r\left(  \dot{\theta}^{2}+\sin^{2}\theta\text{ }\dot{\phi}^{2}\right)
+\frac{qP}{\Omega^{2}}\left[  \cos\theta\left(  \frac{6h^{2}\sin^{2}\theta
}{r^{3}}\right)  +O\left(  \frac{h^{4}}{r^{5}}\right)  \right]  \text{ }%
\dot{\phi},\\
\frac{d}{d\tau}\left(  r^{2}\Omega^{2}\dot{\theta}\right)  =\frac{\dot{t}%
^{2}\Omega_{\theta}}{\Omega^{2}}+\Omega\Omega_{\theta}\left[  \dot{r}%
^{2}+r^{2}\left(  \dot{\theta}^{2}+\sin^{2}\theta\text{ }\dot{\phi}%
^{2}\right)  \right]  +\Omega^{2}r^{2}\left(  \cos\theta\sin\theta\text{ }%
\dot{\phi}^{2}\right)  -\nonumber\\
\frac{q\epsilon\Omega_{\theta}}{\Omega^{2}}\dot{t}+qP\left[  -\sin
\theta\left(  2+\frac{3h^{2}}{r^{2}}\left(  3\cos^{2}\theta-1\right)  \right)
+O\left(  \frac{h}{r}\right)  ^{4}\right]  \text{ }\dot{\phi}.
\end{gather}
From (63-66) it follows that%
\begin{gather}
\text{ }\dot{\phi}\tilde{=}\frac{\left(  \beta_{0}-2qP\cos\theta\right)
}{\Omega^{2}r^{2}\sin^{2}\theta}+O\left(  \frac{\sqrt{h}}{r}\right)  ^{4},\\
\frac{d}{d\tau}\left(  r^{2}\Omega^{2}\dot{\theta}\right)  \tilde{=}%
\Omega_{\theta}\left(  \left[  q\epsilon-\alpha_{0}\Omega-\frac{q\epsilon
}{\Omega}\right]  \left(  q\epsilon-\alpha_{0}\Omega\right)  +\Omega\left[
\dot{r}^{2}+r^{2}\left(  \dot{\theta}^{2}+\sin^{2}\theta\text{ }\dot{\phi}%
^{2}\right)  \right]  \right)  +\nonumber\\
\frac{\left(  qP\right)  ^{2}}{\Omega^{2}r^{2}\sin^{3}\theta}\left(  \beta
\cos\theta-2-\frac{3h^{2}\sin^{2}\theta\cos2\theta}{r^{2}}\right)  \text{
}\left(  \beta-2\cos\theta+\frac{3h^{2}\cos\theta\sin^{2}\theta}{r^{2}%
}\right)  ,
\end{gather}%
\begin{align}
\Omega_{\theta}  &  =-6\frac{mh^{2}\sin\theta\cos\theta}{r^{3}}+O\left(
\frac{h}{r}\right)  ^{4},\\
\Omega_{r}  &  =-\frac{2m}{r^{2}}-\frac{3mh^{2}\left(  3\cos^{2}%
\theta-1\right)  }{r^{4}}+O\left(  \frac{h^{4}}{r^{5}}\right)
\end{align}
so that the latter expressions satisfy the integrability condition,
$\Omega_{\theta r}=\Omega_{r\theta}$ within the range of approximation. We
choose now, similar to the first case of single black hole case, the
particular angles%
\begin{align}
\dot{\theta}  &  =0\rightarrow\theta=\frac{\pi}{2},\\
\dot{\phi}  &  =0\rightarrow\phi=\phi_{0},\nonumber
\end{align}
which give%
\begin{align}
\Omega_{\theta}  &  \simeq0,\\
\Omega_{r}  &  \simeq-\frac{2m}{r^{2}}+\frac{3mh^{2}}{r^{4}}\nonumber\\
\Omega &  \simeq1+\frac{2m}{r}-\frac{mh^{2}}{r^{3}}.\nonumber
\end{align}
We see that the $\phi$ and $\theta$ parts of the equations are trivially
satisfied (by considering the approximation up to the third order) and the two
remaining equations, i.e., $r$ and $t$ parts reduce to the same set of
differential equations which were solved in the previous section, i.e.,
\begin{gather}
-\frac{\dot{t}}{\Omega^{2}}+\frac{q\epsilon}{\Omega}\simeq\alpha_{0},\\
\beta_{0}\simeq0\nonumber\\
\left(  \ddot{r}+\frac{\Omega_{r}}{\Omega}\dot{r}^{2}\right)  \simeq-\left(
q\epsilon-\alpha_{0}\Omega\right)  \frac{\Omega_{r}\alpha_{0}}{\Omega^{2}%
}.\nonumber
\end{gather}
It should be noted also that here the problem yields a different solution,
because the metric function is different. Another special choice of interest
to be considered here is given by%
\begin{align}
\dot{\theta}  &  =0\rightarrow\theta=0,\\
\dot{\phi}  &  =0,\nonumber
\end{align}
which implies%
\begin{align}
\Omega_{\theta}  &  \simeq0,\\
\Omega_{r}  &  \simeq-\frac{2m}{r^{2}}-\frac{6mh^{2}}{r^{4}}\nonumber\\
\Omega &  \simeq1+\frac{2m}{r}+\frac{2mh^{2}}{r^{3}}.\nonumber
\end{align}%
\begin{gather}
-\frac{\dot{t}}{\Omega^{2}}+\frac{q\epsilon}{\Omega}=\alpha_{0},\\
2qP=\beta_{0},\nonumber\\
\left(  \ddot{r}+\frac{\Omega_{r}}{\Omega}\dot{r}^{2}\right)  =-\left(
q\epsilon-\alpha_{0}\Omega\right)  \frac{\Omega_{r}\alpha_{0}}{\Omega^{2}%
}.\nonumber
\end{gather}
These equations also make almost same set of equations as before. Let us add
that generalization to multi-coaxial black hole case (say, along the z-axis)
can be treated more appropriately in the cylindrical polar coordinates. In
these coordinates the electro magnetic potential ansatz takes the form
\begin{equation}
\mathbf{A}=\frac{\epsilon}{\Omega}dt+%
{\textstyle\sum\limits_{i}}
\frac{P_{i}\left(  z-z_{i}\right)  }{\sqrt{\rho^{2}+\left(  z-z_{i}\right)
^{2}}}d\phi
\end{equation}
with
\begin{equation}
\Omega=1+%
{\textstyle\sum\limits_{i}}
\frac{m_{i}}{\sqrt{\rho^{2}+\left(  z-z_{i}\right)  ^{2}}}%
\end{equation}
and the constraint reads as before, namely%
\begin{align*}
\epsilon^{2}+\frac{1}{\lambda^{2}}  &  =1,\\
m_{i}  &  =\left\vert \lambda P_{i}\right\vert =\frac{Q_{i}}{\epsilon}.
\end{align*}
This describes an infinite array of MP black holes, each at $z=z_{i}$, with
coupled electric and magnetic charges, and the line element is given by (1).

\section{Conclusion}

We extend the electrically charged MP black holes to the dyonic case which
possesses both electric ($Q_{i}$) and magnetic ($P_{i}$) charges.
Superposition principle provides us multi-black holes where the mass ($m_{i}$)
of each black hole satisfies $P_{i}^{2}+Q_{i}^{2}=m_{i}^{2}$. The charges are
scaled by a parameter $\epsilon$ ($0\leq\epsilon\leq1$) which regulates the
effective charges of both types. Under such restriction only we were able to
obtain such dyonic solutions. In order to find the interior charge content of
the black hole we provide a detailed analysis of geodesics. Exact particular
integrals are available in some cases but for the general treatment we resort
to the numerical integration and two-dimensional plots. The orbit equation
reveals also the hovering of a test charge around a dyonic black hole. By a
detailed analysis it seems possible that we may identify the charge
constituent of a MP black hole. In a more heuristic argument a magnetically
charged black hole may be identified as a magnetic monopole which, so far has
not been detected in our observable universe. As a final remark we wish to add
that with the inclusion of time in the metric a' la \cite{4} collision problem
of magnetic MP black holes can be investigated.

\textbf{Appendix:}

Regarding the MP line element, the non-zero energy momentum tensor and
Einstein's tensor components are%

\begin{align}
T_{t}^{t}  &  =-\left(  F_{tx}^{2}+F_{ty}^{2}+F_{tz}^{2}\right)  -\frac
{1}{\Omega^{4}}\left[  \left(  F_{xy}^{2}+F_{xz}^{2}+F_{yz}^{2}\right)
\right]  ,\tag{1a}\\
T_{x}^{x}  &  =\left(  -F_{tx}^{2}+F_{ty}^{2}+F_{tz}^{2}\right)  +\frac
{1}{\Omega^{4}}\left(  F_{xy}^{2}+F_{xz}^{2}-F_{yz}^{2}\right)  ,\nonumber\\
T_{y}^{y}  &  =\left(  F_{tx}^{2}-F_{ty}^{2}+F_{tz}^{2}\right)  +\frac
{1}{\Omega^{4}}\left(  F_{xy}^{2}-F_{xz}^{2}+F_{yz}^{2}\right)  ,\nonumber\\
T_{z}^{z}  &  =\left(  F_{tx}^{2}+F_{ty}^{2}-F_{tz}^{2}\right)  +\frac
{1}{\Omega^{4}}\left(  -F_{xy}^{2}+F_{xz}^{2}+F_{yz}^{2}\right)  ,\nonumber\\
T_{x}^{y}  &  =T_{y}^{x}=-2F_{tx}F_{ty}+\frac{2}{\Omega^{4}}\left(
F_{xz}F_{yz}\right)  ,\nonumber\\
T_{x}^{z}  &  =T_{z}^{x}=-2F_{tx}F_{tz}+\frac{2}{\Omega^{4}}\left(
F_{xy}F_{zy}\right)  ,\nonumber\\
T_{y}^{z}  &  =T_{z}^{y}=-2F_{ty}F_{tz}+\frac{2}{\Omega^{4}}\left(
F_{yx}F_{zx}\right)  .\nonumber
\end{align}

\begin{align}
G_{t}^{t}  &  =\frac{1}{\Omega^{4}}\left(  2\Omega\nabla^{2}\Omega-\left(
\mathbf{\nabla}\Omega\right)  ^{2}-3\Omega^{4}\Omega_{t}^{2}\right)
,\tag{1b}\\
G_{x}^{x}  &  =\frac{\Omega_{y}^{2}+\Omega_{z}^{2}-\Omega_{x}^{2}-2\Omega
^{5}\Omega_{tt}}{\Omega^{4}},\nonumber\\
G_{y}^{y}  &  =\frac{\Omega_{x}^{2}+\Omega_{z}^{2}-\Omega_{y}^{2}-2\Omega
^{5}\Omega_{tt}}{\Omega^{4}},\nonumber\\
G_{z}^{z}  &  =\frac{\Omega_{x}^{2}+\Omega_{y}^{2}-\Omega_{z}^{2}-2\Omega
^{5}\Omega_{tt}}{\Omega^{4}},\nonumber\\
G_{x}^{y}  &  =G_{y}^{x}=-2\frac{\Omega_{x}\Omega_{y}}{\Omega^{4}},\nonumber\\
G_{x}^{z}  &  =G_{z}^{x}=-2\frac{\Omega_{x}\Omega_{z}}{\Omega^{4}},\nonumber\\
G_{y}^{z}  &  =G_{z}^{y}=-2\frac{\Omega_{y}\Omega_{z}}{\Omega^{4}},\nonumber\\
G_{t}^{i}  &  =-2\frac{\Omega_{it}}{\Omega^{3}},\text{ }G_{i}^{t}%
=2\Omega\Omega_{it}\nonumber
\end{align}

\begin{align}
T_{t}^{t}  &  =-\left(  \epsilon^{2}+\frac{1}{\lambda^{2}}\right)
\frac{\left(  \Omega_{x}^{2}+\Omega_{y}^{2}+\Omega_{z}^{2}\right)  }%
{\Omega^{4}},\tag{1c}\\
T_{x}^{x}  &  =\left(  \epsilon^{2}+\frac{1}{\lambda^{2}}\right)
\frac{\left(  -\Omega_{x}^{2}+\Omega_{y}^{2}+\Omega_{z}^{2}\right)  }%
{\Omega^{4}},\nonumber\\
T_{y}^{y}  &  =\left(  \epsilon^{2}+\frac{1}{\lambda^{2}}\right)
\frac{\left(  \Omega_{x}^{2}-\Omega_{y}^{2}+\Omega_{z}^{2}\right)  }%
{\Omega^{4}},\nonumber\\
T_{z}^{z}  &  =\left(  \epsilon^{2}+\frac{1}{\lambda^{2}}\right)
\frac{\left(  \Omega_{x}^{2}+\Omega_{y}^{2}-\Omega_{z}^{2}\right)  }%
{\Omega^{4}},\nonumber\\
T_{x}^{y}  &  =T_{y}^{x}=-\left(  \epsilon^{2}+\frac{1}{\lambda^{2}}\right)
\frac{2\Omega_{x}\Omega_{y}}{\Omega^{4}},\nonumber\\
T_{x}^{z}  &  =T_{z}^{x}=-\left(  \epsilon^{2}+\frac{1}{\lambda^{2}}\right)
\frac{2\Omega_{x}\Omega_{z}}{\Omega^{4}},\nonumber\\
T_{y}^{z}  &  =T_{z}^{y}=-\left(  \epsilon^{2}+\frac{1}{\lambda^{2}}\right)
\frac{2\Omega_{y}\Omega_{z}}{\Omega^{4}},\nonumber
\end{align}

\textbf{Figure captions:}

Fig 1a: Freely falling charged particle into the dyonic black hole as a
function of proper time. In a finite proper time the particle reaches the
horizon, as expected. With the magnetic charge on the black hole, the test
particle plunges into the black hole in a shorter proper time. The infall gets
delayed for a weaker magnetic charge.

Fig 1b: The free fall motion of a test charge is observed from a far distance.
It takes an infinite coordinate time to reach the horizon and the magnetic
charge has little effect in the process.

Fig 2: The oscillatory motion of a test charge around a dyonic black hole.
$x\left(  \phi\right)  $($=r\left(  \phi\right)  $) is plotted versus the
azimuthal angle. Our boundary conditions are such that $x\left(
\phi=0\right)  =1$ and $\left.  \frac{dx}{d\phi}\right\vert _{\phi=0}=0,$ the
rest is determined by the differential equation of orbit. An exact solution,
which is not at our disposal, should definitely reveal much more than our
numerical analysis. The effect of the magnetic charge on the behavior of the
test particle is evidently visible.


\begin{thebibliography}{9}                                                                                                %


\bibitem {1}S. D. Majumdar, Phys. Rev. \textbf{72}, 390 (1947).

\bibitem {2}A. Papapetrou, Proc. R. Ir. Acad., A Math. Phys. Sci.
\textbf{A51}, 191 (1947).

\bibitem {3}J. B. Hartle and S. W. Hawking, Commun. Math. phys. \textbf{26},
87 (1972).

\bibitem {4}D. Kastor and J. H. Traschen, Phys. Rev. D \textbf{47}, 5370 (1993).

\bibitem {5}D. R. Brill, G. T. Horowitz, D. Kastor and J. H. Traschen, Phys.
Rev. D \textbf{49}, 840 (1994).

\bibitem {6}G. W. Gibbons, H. L\"{u} and C. N. Pope Phys. Rev. Lett.
\textbf{94}, 131602 (2005).

\bibitem {7}W. Israel and G. A. Wilson, J. Math. Phys. \textbf{13}, 865 (1972).

\bibitem {8}R.M. Corless, G.H. Gonnet, D.E.G. Hare, D.J. Jeffrey, D.E. Knuth.
Adv. Comput. Math. \textbf{5}, 329 (1996) .
\end{thebibliography}
\end{document}